\documentstyle[12pt,epsf]{article}

\def\bnabla{\mbox{\boldmath$\nabla$}}

\def\bp{{\bf p}}
\def\br{{\bf r}}

\def\bz{{\bf z}}

\def\bs{{\bf s}}

\def\bL{{\bf L}}

\def\bB{{\bf B}}

\def\bA{{\bf A}}

\newcommand{\be}{\begin{equation}}
\newcommand{\ee}{\end{equation}}
\newcommand{\bea}{\begin{eqnarray}}
\newcommand{\eea}{\end{eqnarray}}

\newcommand{\lan}{\langle}
\newcommand{\ran}{\rangle}

\newcommand{\zp}{\dot{z}_{\perp}}

\newcommand{\p}{\partial}
\begin{document}

\vskip 2cm
\begin{center}
\Large
{\bf  The minimum area, the flux tube, } \\
{\bf and Thomas precession } \\
\vskip 0.5cm
\large
 Ken Williams \\
{\tiny \em General Delivery /  Ojai, CA. 93023 } \\
\end{center}
\thispagestyle{empty}
\vskip 0.7cm

\begin{abstract}

  Quark confinement in Buchm\"{u}ller's picture of a rotating flux
  tube is reconsidered in the context of a minimum area evaluation of
  the Wilson loop. The question is asked whether resulting spin
  independent dynamics are consistent with Thomas precessional spin
  dependence leading from electric confinement. The answer appears to
  be in the negative; self-consistency of the picture found in the
  literature is examined and explained in simple classical terms, with
  illustrations.

\end{abstract}

\newpage

\section{introduction}

The problem of color confinement has been with us for a long time.
When quarks and gluons were first proposed as the fundamental
constituents of the hadron there was little understanding of the
mechanism responsible for the absence of free colored states.
Wilson's observation that the amplitude of a gauge invariant meson
decays exponentially in the sum over areas of world sheets swept out
by paths connecting the $ Q\bar{Q} $ world lines led to his so called
adiabatic minimum area law whose driving force is gauge invariance.
Based on this invariance Eichten and Feinberg a few years later
derived the most general $ Q\bar{Q} $ interaction Hamiltonian
\cite{ef} valid to $O(v^2) $.

By these and other results \cite{con,gromes} there has gradually
emerged the view that the string or tube corresponding to the world
sheet at a given time slice is characterized by a purely color
electric flux.  Such was the picture of things proposed by
Buchm\"{u}ller \cite{buch} to answer the question how short and long
range vector couplings might together yield by cancellation an
observed small spin-orbit splitting: Long range color magnetic fields
in the co-moving rest frame are set to zero resulting in kinematic
Thomas precessional spin dependence only.

It was no more than a decade ago that spin independent relativistic
corrections too were calculated in the Minimal Area Law approximation
(MAL) of the Wilson loop \cite{bram} and shown shortly after to be
equivalent to those of a rotating flux tube \cite{oow}. Spin
dependence so derived agrees with predictions from electric
confinement. In the meantime explicit flux tube dynamics has been
shown to follow beginning at the other end from an electric
confinement ansatz \cite{ol} , so that in today's literature
self-consistency of the picture is fairly taken for granted
\cite{isgur}.

The hope here is to shed a little light on the conditions necessary
for that self-consistency. This by way of example.  A review of the
procedure in the MAL of ref\cite{bram} by which spin dependence of the
interaction Hamiltonian is derived is presented in section 2.  Purely
Thomas precessional spin-orbit is shown to follow in the end from a
mathematical inconsistency whose physical interpretation is,
fortunately, transparent. Section 3 considers the same problem as
treated in ref\cite{ol} beginning from the given spin dependence of
electric confinement. Spin independent flux tube dynamics in this case
follows from the inverted inconsistency of the previous case.

In reviewing each approach the scheme here is to 1) clearly
demonstrate the respective mathematical inconsistencies leading to
Thomas precessional spin dependence with rotational flux tube spin
independence, 2) re-state these inconsistencies in terms more
physically transparent, and 3) propose a mathematically consistent
approach to the problem.

\section{Thomas precession from the minimum area}

Beginning from a gauge invariant 4-point function the $Q\bar{Q} $
interaction Hamiltonian of ref\cite{bram} is derived to $O(\dot{z}^2)$
in terms of the Wilson loop and its expectation values of the field
strength tensor
\bea
V &=& V_{so} + V_{si} \\
V_{so} &=& \frac{1}{2} \epsilon_{ijk} s^i  \lan\lan \dot{z}^j F_{0k} + 
 F_{jk} \ran\ran \label{two} \\
\int dt V_{si} &=& i \ln W = i\ln \frac{1}{3} \lan tr P
\exp( ig\oint dt ( A_0 - \dot{z}^i A^i ) ) \ran
\eea
where for comparison with results from \cite{ol} the antiquark mass is
taken to infinity, leaving z as the quark coordinate. The Darwin term,
irrelevant to the present discussion, is suppressed. In the MAL the
Wilson loop is approximated by a Nambu-Goto string action
\bea
{\cal S}[u] &=& a \int dt \int^{1}_{0} ds [ (\dot{u} \cdot u^{\prime } )^2
- \dot{u}^2 u^{\prime 2} ]^{1/2} \nonumber
 \\ &\equiv & a\int dt \int^{1}_{0} ds S
\eea
constrained by Euler-Lagrange minimizing relations
\bea
\frac{\p}{\p s} \left( \frac{\p S}{\p u^{\prime i } } \right) +
\frac{\p}{\p t} \left( \frac{\p S}{\p \dot{u}^i } \right) &=& 0
\eea
which to $O(\dot{z}^2)$ gives effectively a straight-line
approximation for the area
\bea
u_{min}^i &\approx & s z^i -  s(1-s^2) \zeta^i/6 \, ;
 \quad \zeta \sim O(\dot{z}^2 )  \nonumber \\
u_0 &=& z_0 \,.
\eea
where $ u_\mu $ is a point on an area bounded by the loop.  The MAL is
therefore
\bea
\imath \ln W[z_\mu ] & = &  {\cal S }[u_\mu ]|_{u=u_{min}} \label{mal}
 \\
i\ln \frac{1}{3} \lan tr P \exp( ig\oint dt ( A_0 - \dot{z}^i A^i ) ) \ran &=&
a \int dt \int^{1}_{0} du_{min}
 ( 1 - \dot{u}^{2}_{ min 
\perp } )^{1/2} \nonumber \\
& \simeq & \int dt \int^{1}_{0} ds
az (1  -  s^2 \zp^2/2)  \nonumber \\
&=& \int dt (az  - a z \zp^2/6)
\eea
immediately yielding from (3)
\bea
V_{si} &\simeq & az - \frac{aL^2}{6m^2 z}
\eea
which from classical considerations has been shown to be consistent
with the dynamics of a rotating tube of constant energy
density\cite{oow}.

What remains for determination of spin dependent contributions,
eq.(\ref{two}), is the evaluation of the field strength tensor Wilson
loop expectation values.  These are obtained in \cite{bram} via the
functional variation
\bea
\delta \imath \ln W[z_\mu ] = ( \delta {\cal S }[u_\mu ] )|_{u=u_{min}}
\label{var1} 
\eea
as distinct from a variation on the above MAL
\bea
\delta \imath \ln W[z_\mu ] = \delta ( {\cal S }[u_\mu ]|_{u=u_{min}} ).
\label{var2}
\eea
The resulting $ O(\dot{z}^2) $ Lorentz force is
\bea
\lan \lan F_{0i} + \dot{z}^j F_{ji} \ran \ran &\simeq & a [
\hat{z}^i ( 1 +
\frac{1}{2}\zp^2 ) + \dot{z}^j (\hat{z}^j \dot{z}^i - \hat{z}^i \dot{z}^j )]  
 \label{twelve}
\eea
from which the obvious identifications are assumed
\bea
\lan \lan F_{0i} \ran \ran &= & a
 \hat{z}^i ( 1 + \frac{1}{2}\zp^2 ) 
\nonumber \\
\lan \lan F_{ji} \ran \ran &= & a
(\hat{z}^j \dot{z}^i - \hat{z}^i
\dot{z}^j )
\eea
yielding the expected spin dependent Thomas precession of electric
confinement
\bea
V_{so} &\simeq & - \frac{a}{2m^2 z} \bs \cdot \bL \, .
\eea
What is telling about this result is that the same is obtained by
assuming the flux tube orientation fixed ( $\hat{\bz} = const $) and
its velocity externally prescribed $(\dot{z } = f_{ext}(t))$, so that
instead of rotational motion about the center of momentum as depicted
in fig(1) the classical physics is that of a tube with its center of
momentum in rectilinear motion, fig(2).

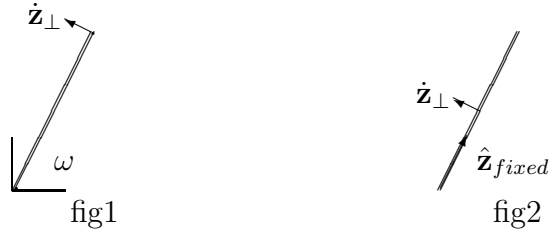
\begin{picture}(300,100)
\put(78,12){\line(0,1){20}}
\put(78,12){\line(1,0){20}}
\put(78,12){\line(1,2){30}}
\put(90,20){ $ \omega $ }
\put(100,0){fig1}
\put(79,12){\line(1,2){30}}
\put(78,12){.}
\put(107,71){.}
\put(105,73){\vector(-2,1){7}}
\put(108,72){\line(-2,1){10}}
\put(80,75){ $ \dot{\bz}_\perp $ }

\put(240,12){\line(1,2){30}}
\put(239,12){\line(1,2){30}}
\put(250,20){ $ \hat{\bz}_{fixed} $ }
\put(240,13){\vector(1,2){10}}
\put(252,43){\vector(-2,1){7}}
\put(255,42){\line(-2,1){10}}
\put(227,45){ $ \dot{\bz}_\perp $ }
\put(260,0){fig2}
\end{picture}

In fact, from the mechanics of constrained systems\cite{gold} ( or
from simple counting of degrees of freedom ) it is clear that
(\ref{var1}) is not a variation of (\ref{mal}), the MAL. Thomas
precessional spin-orbit, i.e. eq.(\ref{twelve}), is obtained from the
MAL variation, variation (\ref{var2}), by 1) taking the tube
velocity to be externally specified, 2) performing the coordinate
variation, and 3) fixing the tube's orientation
\bea
\delta \imath \ln W[z_\mu ] &=& \int dt \delta z^i \lan \lan
F_{0i} + \dot{z}^j F_{ji} \ran \ran \nonumber 
\\ &=& \delta ( {\cal S }[u_\mu
]|_{u=u_{min}} ) \simeq \delta a
\int dt \int^{1}_{0} ds z (1 - s^2
\zp^2 ) \nonumber
\\ & \to & a
 \int dt \delta z^i \frac{\p}{\p z^i} \, z(1 -
\zp^2 ) \nonumber 
\\ & \approx & a
 \int dt \delta z^i [ \hat{z}^i ( 1 +
\frac{1}{2}\zp^2 ) + \dot{z}^j (\hat{z}^j \dot{z}^i - \hat{z}^i
\dot{z}^j ) ]
\eea
so that the straight-line-area velocity for rotation, $
\dot{u}^{i}_\perp =s\dot{z}^{i}_\perp $, where s varies along the flux
tube $ (0 \le s \le 1 ) $ , is replaced by $ \dot{u}^{i}_\perp
=\dot{z}^{i}_\perp $ , thus assigning to each point the same
perpendicular velocity, the resulting physics corresponding to that
illustrated in fig(2) ( When $m_2 $ is finite what happens is even
more transparent. Then, $ \dot{u}^{i}_\perp =s\dot{z}^{i}_{ 1 \perp }
+ (1-s) \dot{z}^{i}_{2 \perp } $, so that fixing the tube orientation,
$ \hat{\br } $ , gives $ \dot{z}^{i}_{ 1 \perp } = \dot{z}^{i}_{ 2
  \perp } $, or, $ \dot{u}^{i}_\perp =\dot{z}^{i}_\perp $ ). The
inconsistency consists in leaving the minimal area evaluation,
(\ref{mal}), for spin independence untransformed.  Going through
however with this transformation on $ V_{si} $ further clarifies that
the classical physics in this case corresponds that of a tube of
constant energy density in rectilinear motion
\bea
V_{si} &\simeq & \bar{m} - 
\frac{1}{2} \bar{m} \zp^2 \, , \quad \bar{m} \equiv a z
\eea
and not that of a rotating tube. 

For the physics appropriate to this bound state problem, that pictured
in fig(1), one obtains the Lorentz force from variation (\ref{var2})
\bea
\lan \lan F_{0i} + \dot{z}^j F_{ji} \ran \ran &\simeq & a [\hat{z}^i ( 1 -
\frac{1}{6}\zp^2 ) + \frac{1}{3}\dot{z}^j (\hat{z}^j \dot{z}^i - \hat{z}^i \dot{z}^j ) +\frac{1}{3} z \ddot{z}^i_\perp ]
 \, . \label{ten}
\eea
At this point one has to exercise some care. The identifications
\bea
\lan \lan F_{0i} \ran \ran &= & a [\hat{z}^i ( 1 - \frac{1}{6}\zp^2 )
+ \frac{1}{3} z \ddot{z}^i_\perp ]
\nonumber \\
\lan \lan F_{ji} \ran \ran &= & \frac{1}{3} a 
(\hat{z}^j \dot{z}^i - \hat{z}^i \dot{z}^j )
\eea
while consistent from the discussion so far are premature; other
assignments are possible. I.e., (\ref{ten}) is insufficient to
determine $ \langle \langle F^{\mu \nu } \rangle \rangle $ ; the MAL
itself provides the additional constraint.

Taylor expanding (\ref{ten}) and (\ref{mal}) as function and
functional, respectively, of $\dot{z}^{i} $ to second order (see
appendix) yields the identifications
\bea 
\lan \lan F_{0i} \ran \ran &= & a [
\hat{z}^i ( 1 + \frac{1}{6}\zp^2 )
+ \frac{1}{3} \hat{z}^j ( z^j \ddot{z}^i_\perp - \dot{z}^j \dot{z}^i_\perp )] 
\nonumber \\ \lan
\lan F_{ji} \ran \ran &= & \frac{2}{3} a 
(\hat{z}^j \dot{z}^i - \hat{z}^i
\dot{z}^j ) \label{res2} 
\eea
with a resulting spin-orbit interaction differing with that
of Thomas precession
\bea
V_{so} &\simeq & - \frac{a}{6m^2 z} \bs \cdot \bL
\eea
though consistent with that of rotational flux tube dynamics
\cite{ovw}.

\section{ flux tube dynamics from electric confinement }

Buchm\"{u}ller's picture of a color electric flux tube in the
co-moving rotating frame of the $Q\bar{Q} $ system immediately obtains
in \cite{ol} Thomas spin-orbit dependence from the $O(\dot{z}^2) $
reduced Salpeter Hamiltonian
\bea
V &=& V_{so} + V_{si} \\
V_{si} &=& A_0 - \dot{z}^i A^i \\
V_{so} &=& \frac{1}{2} \epsilon_{ijk} s^i ( \dot{z}^j F_{0k} +
 F_{jk} ) \\ & \simeq & -\frac{a}{2m^2 z} \bL \cdot \bs
\eea
where
\bea
F_{0i} &=& E^i = -a \hat{z}^i (1-\zp^2)^{-1/2} \nonumber \\
F_{ji} &=& - \epsilon_{jik} B^k = a (\hat{z}^j \dot{z}^i - \hat{z}^i
\dot{z}^j ) (1-\zp^2)^{-1/2} \label{con}
\eea
are found by Lorentz transforming (infinitesimally) from the co-moving
frame\cite{jack}. Spin independent $ V_{si} $ is obtained from the
differential gauge field constraints (\ref{con}). The proposed
solutions are
\bea
A_0 &=& az\int^{1}_{0} ds ( 1-\dot{u}^2_\perp )^{-1/2} \nonumber \\
A^i &=& az \zp^i \int^{1}_{0} ds s^2 ( 1-\dot{u}^2_\perp )^{-1/2}
\label{sol}
\eea
giving
\bea
V_{si} &=& az\int^{1}_{0} ds ( 1-\dot{u}^2_\perp )^{1/2} = 
az\int^{1}_{0} ds ( 1- s^2\dot{z}^2_\perp )^{1/2} \nonumber \\
&\approx & az - \frac{aL^2}{6m^2 z}
\eea
the expected flux tube dynamics. The question is simply whether
solutions (\ref{sol}) indeed satisfy (\ref{con}). This is checked in
\cite{ol} by taking the appropriate derivatives while holding the term
$ \omega \equiv \zp /z $ spatially fixed
\bea
E^i &=& - \nabla^i A_0 \to \nabla^{i}_{|\bz |} A_0 = -a \hat{z}^i (
1- \dot{z}^2_\perp )^{-1/2} \nonumber \\
\bB &=& \bnabla \times \bA \to
\bnabla_{|\bz |} \times \bA = a \hat{\bz } \times \dot{\bz } ( 1-
\dot{z}^2_\perp )^{-1/2} 
\eea
an unfortunate artifact of which is the appearance in the Hamiltonian
of two distinct expressions for the orbital angular momentum operator
- one dependent on spatial orientation
\bea
V_{so} & \simeq & -\frac{a}{2m^2 z} \bL \cdot \bs \, , \quad L = 
z (p^2 - (\hat{\bz} \cdot \bp )^2 )^{1/2}
\eea
and the other independent
\bea
V_{si} &\simeq & az - \frac{aL^2}{6m^2 z} \, , \quad L \simeq m z^2 \omega \, ;
\quad
 \omega \neq \omega (\hat{\bz} ) \, . 
\eea
A clear inconsistency.

On the other hand the functions
\bea
A_0 &=& az (1-\dot{z}^2
)^{-1/2} \nonumber \\ A^i &=& az \dot{z}^i (1-\dot{z}^2 )^{-1/2} 
\eea
yielding 
\bea
V_{si} &=& A_0 - \dot{z}^i A^i = az (1-\dot{z}^2
)^{1/2} \nonumber \\ & \approx & az - \frac{a L^2}{2m^2 z}
\eea
are in fact solutions of (\ref{con}) , although spin independent
angular corrections from these are again those of the flux tube in
rectilinear motion shown in fig(2). One might very well have guessed
from the discussion in section 2 that spin independent rotational flux
tube dynamics is here obtained by the inverse transformation taken
there , i.e., by transforming from linear variables $ z_i $ to
rotational ones $ u_i $
\bea
A_0 - \dot{z}^i
A^i & =& az (1-\dot{z}^2 )^{1/2} \to a \int^{1}_{0} ds u (1-\dot{u}^2
)^{1/2} \nonumber \\ & \approx & az - \frac{a L^2}{6 m^2 z} 
\eea 
with 
\bea
A_0 &=& az (1-\dot{z}^2 )^{-1/2} \to a \int^{1}_{0} ds u
(1-\dot{u}^2 )^{-1/2} \nonumber \\ \dot{z}^i A^i &=& az \dot{z}^2
(1-\dot{z}^2 )^{-1/2} \to a \int^{1}_{0} ds u \dot{u}^2 (1-\dot{u}^2
)^{-1/2} 
\eea 
or
\bea
A^i& = &a \int^{1}_{0} ds u \dot{u}^i
(1-\dot{u}^2 )^{-1/2} = az \zp^i \int^{1}_{0} ds s^2 (
1-\dot{u}^2_\perp )^{-1/2} 
\eea 
the inconsistency here consisting in leaving spin dependent terms for
$ V_{so} $ untransformed.

\section{summary}

In the above discussion electric confinement is shown to be
incompatible with the MAL formulation of rotational flux tube
dynamics. This incompatibility appears in the form of discrepancies
between spin dependent and independent $O(\dot{z}^2)$ corrections to
the interaction Hamiltonian. It is demonstrated that the agreement
brought about in the literature follows from inconsistent treatment of
the flux tube orientation vector $ \hat{\bz} $ ; the MAL and electric
confinement ansatz are connected via the dynamics of a rectilinearly
moving non-rotating flux tube.

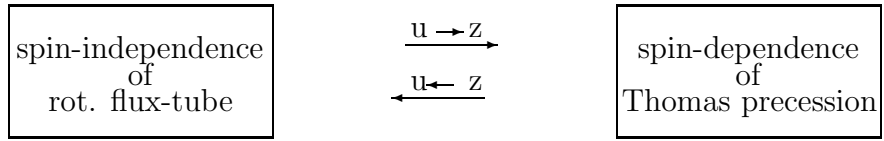
\begin{picture}(300,100)
\put(20,80){\line(0,-1){50}}
\put(20,80){\line(1,0){100}}
\put(20,30){\line(1,0){100}}
\put(120,80){\line(0,-1){50}}

\put(250,80){\line(0,-1){50}}
\put(250,80){\line(1,0){100}}
\put(250,30){\line(1,0){100}}
\put(350,80){\line(0,-1){50}}

\put(170,65){\line(1,0){30}}
\put(200,45){\line(-1,0){30}}
\put(200,65){\vector(1,0){5}}
\put(170,45){\vector(-1,0){5}}
\put(182,70){\line(1,0){5}}
\put(188,50){\line(-1,0){5}}
\put(187,70){\vector(1,0){5}}
\put(183,50){\vector(-1,0){5}}
\put(194,68){z}
\put(172,68){u}
\put(194,48){z}
\put(172,48){u}

\put(23,60){spin-independence}
\put(65,50){of}
\put(35,40){rot. flux-tube}

\put(258,60){spin-dependence}
\put(295,50){of}
\put(252,40){Thomas precession}

\end{picture}

On a purely rational level the conclusion should come as no surprise.
It is well known that the tendency to desire a `` $ - \frac{1}{2} $ ``
numerical factor for $ Q\bar{Q} $ spin-orbit interactions began with
the early successes of the scalar confinement ansatz. One should
simply ask the question whether scalar confinement follows by
necessity from rotational flux tube dynamics.  The same question could
be asked with regard to scalar confinement from Lorentz invariance
\cite{gromes} or covariance. The question has been asked \cite{ken}.
In both cases it is a question of mathematical deduction ( assuming
the terms of the model agreed upon, i.e., that the model is
well-defined ).

The present result is not entirely negative. A consistent approach
beginning from the MAL yielding spin dependent and independent
corrections identical with those of ref\cite{ovw} has been presented
in section 2. Also, beginning from the electric confinement ansatz a
consistent approach again yielding both spin dependent and independent
corrections is maintained in \cite{aow}. Predictions from the two
distinct models themselves remain distinct.

\section{appendix}

From the Taylor function and functional expansions
\bea f( \dot{\bz }) &=& f(0) + \dot{z}^i \int dt^\prime \left(
  \frac{\delta }{ \delta \dot{z}^i } f(\dot{\bz}^\prime ) \right)_{0}
\nonumber 
\\ && + \frac{1}{2} \dot{z}^i \dot{z}^j \int \int dt^\prime dt^{\prime
  \prime } \left( \frac{\delta }{ \delta \dot{z}^i } \frac{\delta }{
    \delta \dot{z}^{\prime j }} f(\dot{\bz}^{\prime \prime })
  \right)_{0} \, + h.o. \\ F[ \dot{\bz }] &=& F[0] + \int dt \dot{z}^i
  \left( \frac{\delta }{ \delta \dot{z}^i } F[\dot{\bz}^\prime ]
  \right)_{0} \nonumber 
\\ && + \frac{1}{2} \int \int dt dt^\prime \dot{z}^i
  \dot{z}^j \left( \frac{\delta }{ \delta \dot{z}^i } \frac{\delta }{
      \delta \dot{z}^{\prime j }} F[\dot{\bz}^{\prime \prime }]
    \right)_{0} \, + h.o.  \eea
    the spatial field tensor Wilson loop expectation value to
first order is
\bea
\lan \lan F^{ij} \ran \ran &=& \lan \lan F^{ij} \ran \ran_{0} +
\dot{z}^k \int dt^\prime \left( \frac{\delta }{ \delta  \dot{z}^k } 
\lan \lan F^{ij} \ran \ran^\prime \right)_0 \nonumber \\
&=& 0 + \dot{z}^k \int dt^\prime \left( \frac{\delta }{ \delta  \dot{z}^k } 
\lan \lan \p_i A^j \ran \ran^\prime  - \frac{\delta }{ \delta  \dot{z}^k } 
\lan \lan \p_j A^i \ran \ran^\prime \right)_0  \, . 
\eea
To evaluate the rhs the MAL, (\ref{mal}), is expanded
\bea
\imath \ln W  &=& \imath \ln W_0 -
\int dt \dot{z}^i \lan \lan A^i \ran \ran_0 \nonumber \\ & &
- \frac{1}{2} \int dt dt^\prime \dot{z}^i \dot{z}^j \left(   \frac{\delta }{ \delta  \dot{z}^i } 
\lan \lan \p_i A^j \ran \ran^\prime  \right)_0 \, + h.o. \nonumber  \\
&=& \int dt [ az - \dot{z}^i ( \frac{1}{6} a z \zp^i ) ]
\eea
giving
\bea
\dot{z}^i \lan \lan A^i \ran \ran_0 &=& 0 \nonumber \\
\dot{z}^k \int dt^\prime \left( \frac{\delta }{ \delta  \dot{z}^k } 
\lan \lan  A^j \ran \ran^\prime \right)_0 &=& \frac{1}{3} a z \zp^j
\label{res1}
\eea
as sufficient conditions, where
\bea
\frac{\delta }{ \delta  \dot{z}^j } \imath \ln  W &=&
\frac{\delta }{ \delta  \dot{z}^j } \imath \ln \frac{1}{3}
 \lan  tr P \exp[ \imath g
\int dt^\prime ( A_0 -\dot{z}^i A^i ) ]  \ran \nonumber \\
&=& - \lan \lan A^j \ran \ran
\eea
has been used. Then the coordinate derivative of the lhs of
(\ref{res1}) is
\bea
& & \frac{\p}{\p z^{\prime \prime i }} \dot{z}^k \int dt^\prime \left( \frac{\delta }{ \delta  \dot{z}^k } 
\lan \lan  A^j \ran \ran^\prime \right)_0
\to  \int dt^{\prime \prime } \frac{\delta}
{\delta z^{\prime \prime i }} \dot{z}^k \int dt^\prime \left( \frac{\delta }{ \delta  \dot{z}^k } 
\lan \lan  A^j \ran \ran^\prime \right)_0 \nonumber \\
&=& \int dt^{\prime \prime } \dot{z}^k  \left( \frac{\delta }{ \delta  \dot{z}^k } 
\lan \lan \p_i A^j \ran \ran^{\prime \prime }
- \frac{\delta }{ \delta  \dot{z}^k } \int dt^\prime
 [  \lan \lan  A^j \ran \ran^\prime  \lan \lan  \dot{z}^\mu 
 F_{\mu i } \ran \ran^\prime  - \lan \lan  A^j  \dot{z}^\mu 
 F_{\mu i } \ran \ran^\prime  ]
\right)_0 \nonumber 
\\  &\to & \int dt^{\prime } \dot{z}^k  \left( \frac{\delta }{ \delta  \dot{z}^k } 
\lan \lan \p_i A^j \ran \ran^{\prime  } \right)_0
\eea
from its function as opposed to functional character. Then
\bea
\dot{z}^k \int dt^\prime \left( \frac{\delta }{ \delta  \dot{z}^k } 
\lan \lan \p_i A^j \ran \ran^\prime \right)_0
&=& \frac{1}{3} a [ \hat{z}^i \dot{z}^j - \hat{z}^j \dot{z}^i +
 \hat{\bz} \cdot \dot{\bz} ( \hat{z}^i \hat{z}^j - \delta_{ij} ) ]
\eea
yielding
\bea
\lan \lan F^{ij} \ran \ran &=& \frac{2}{3} a ( \hat{z}^i \dot{z}^j - \hat{z}^j \dot{z}^i )
\eea
equation (\ref{res2}).

\end{document}